\documentstyle{article}

\newcommand{\ba}{\begin{eqnarray}}
\newcommand{\ea}{\end{eqnarray}}
\input{psfig}

\begin{document}
\pagestyle{plain}

\title{Algebraic treatment of the hypercoulomb problem}
\author{R. Bijker\\
Instituto de Ciencias Nucleares, \\ 
Universidad Nacional Aut\'onoma de M\'exico, \\ 
Apartado Postal 70-543, 04510 M\'exico, D.F., M\'exico 
\and 
F. Iachello\\
Center for Theoretical Physics, Sloane Laboratory,\\
Yale University, New Haven, CT 06520-8120, U.S.A.
\and 
E. Santopinto\\
Dipartimento di Fisica dell'Universit\`a di Genova,\\
Istituto Nazionale di Fisica Nucleare, Sezione di Genova, \\
via Dodecaneso 33, 16164 Genova, Italy}
\date{}
\maketitle

\begin{abstract}
A completely algebraic treatment of the 
six-dimensional hypercoulomb problem is discussed in terms of an 
oscillator realization of the dynamical algebra of $SO(7,2)$.  
Closed expressions are derived for the energy spectrum and 
form factors. 
\end{abstract} 

\clearpage

\section{Introduction}

Recently there has been renewed interest in the study of 
the three-body problem, mostly arising from constituent quark 
models of baryons in terms of three valence quarks.  
In these problems, the hypercentral approximation \cite{bafa}, which 
means using a hypercentral potential, provides often a good approximation 
to the actual situation \cite{FGPS,SIG}. Among the hypercentral
potentials, two play a special role: the six-dimensional harmonic 
oscillator and the six-dimensional coulomb potential, since both 
problems are exactly solvable. 
In this article, we present a completely algebraic solution of the 
hypercoulomb problem. The method is based on the use of $SO(7,2)$ as a 
dynamical group of the six-dimensional hypercoulomb potential 
\cite{Barut}. It is a generalization of $SO(4,2)$ dynamical 
group of the ordinary hydrogen atom \cite{Hatom}.

In addition to obtaining a series of mathematical results 
associated with this rather complicated algebra, we briefly
point out that the hypercoulomb potential can be used as a good 
approximation to actual physical situations. One occurs in atomic 
physics and it was mentioned long ago in \cite{Fabre}. 
The authors showed that the two-body coulomb potential in the hypercentral 
approximation, that is the hypercoulomb potential,
provides a good description of the lower atomic states. 
Another application occurs in hadronic physics, in the constituent 
quark model of baryons. Due to the extraordinary difficulty to calculate 
physical quantities from QCD (in the nonperturbative regime) one has 
to rely on models, like constituent quark models. The short range 
behaviour is dominated by coulomb potentials as seen from the 
nonrelativistic reduction of the one-gluon exchange diagram \cite{OGE} 
and the long range one by a linear potential as from results of lattice 
calculations. So one considers the potential dominated by a coulomb-like 
plus a linear confining term \cite{FGPS,cornell,tax}. 
The coulomb like term provides a good approximation to the low-lying  
states and the effects of the linear confining terms plus other small 
contributions can be taken into account in perturbation theory as shown in 
\cite{SIG}. This application is discussed in Section 8.  
Particular emphasis is paid to form factors which are the quantities 
of direct experimental interest, but which in general are very 
difficult to calculate.  

\section{The hypercoulomb potential} 

We consider the hypercoulomb potential in six dimensions
\ba
H &=& \frac{p^2}{2\mu} - \frac{\tau}{r} ~, \label{hamilt}
\ea
with $p^2 = \sum_{j=1}^6 p_j^2$ and $r^2 = \sum_{j=1}^6 r_j^2$. 
The energy eigenvalues and the degeneracy of the eigenstates  
can be obtained by studying the dynamical group $SO(7,2)$ \cite{Barut}. 
The generators of the Lie algebra of $SO(7,2)$ can be realized 
in coordinate space by 
\ba
L_{jk} &=& r_j p_k - r_k p_j ~,
\nonumber\\ 
L_{j7} &=& \frac{1}{2} r_j p^2 - p_j \sum_{k=1}^6 r_k p_k 
+ \frac{3}{2} i p_j - \frac{1}{2} r_j ~,
\nonumber\\
L_{j8} &=& \frac{1}{2} r_j p^2 - p_j \sum_{k=1}^6 r_k p_k  
+ \frac{3}{2} i p_j + \frac{1}{2} r_j ~,
\nonumber\\ 
L_{j9} &=& r p_j ~, 
\nonumber\\
L_{78} &=& \sum_{k=1}^6 r_k p_k - \frac{5}{2} i ~,
\nonumber\\ 
L_{79} &=& \frac{1}{2} (r p^2 - r) ~,
\nonumber\\ 
L_{89} &=& \frac{1}{2} (r p^2 + r) ~,
\label{gen1}
\ea 
with $j,k=1,\ldots,6$. These generators are antisymmetric 
$L_{ij}=-L_{ji}$, and satisfy the commutation relations 
\ba
\left[ L_{ij},L_{kl} \right] &=& -i \, ( g_{ik} \, L_{jl} 
+ g_{jl} \, L_{ik} - g_{il} \, L_{jk} - g_{jk} \, L_{il} ) ~,
\ea
with
\ba
g_{ij} &=& \left\{ \begin{array}{ll} 
- \delta_{ij} & \mbox{  for } j=1,\ldots,7 ~, \\
+ \delta_{ij} & \mbox{  for } j=8,9 ~. \end{array} \right. 
\ea
The $SO(7,2)$ algebra combines the $SO(2,1)$ spectrum generating 
algebra of the hypercoulomb problem and its $SO(7)$ degeneracy group 
into a single algebraic structure \cite{Barut}. The corresponding 
subgroup chains are 
\ba
SO(7,2) \supset \left\{ \begin{array} {ccc}
SO(6) &\otimes& SO(2,1) \\ \gamma &,& q \\
& & \\ SO(7) &\otimes& SO(2) \\ \omega &,& q_0 
\end{array} \right\} \supset SO(6) \otimes SO(2) ~. \label{lattice}
\ea
The eigenstates of the hypercoulomb problem belong to a single  
infinite dimensional representation of $SO(7,2)$ \cite{Barut}, 
which can be decomposed into the irreducible representations of its 
subgroups to provide a complete set of basis states.  
The generators and the quadratic Casimir operators of the groups 
appearing in the group lattice of Eq.~(\ref{lattice}) are given by 
\ba
SO(7)   \, : & \hspace{1cm} L_{jk} \, (j,k=1,\ldots,7) 
& \hspace{1cm} \Lambda_7^2 \;=\; \frac{1}{2} \sum_{j,k=1}^7 L_{jk}^2 ~, 
\nonumber\\
SO(6)   \, : & \hspace{1cm} L_{jk} \, (j,k=1,\ldots,6) 
& \hspace{1cm} \Lambda_6^2 \;=\; \frac{1}{2} \sum_{j,k=1}^6 L_{jk}^2 ~, 
\nonumber\\
SO(2,1) \, : & \hspace{1cm} T_1=L_{79} ~, \, T_2=L_{78} ~, \, T_3=L_{89} 
& \hspace{1cm} T^2 \;=\; T_3^2 - T_1^2 - T_2^2 ~, 
\nonumber\\
SO(2)   \, : & \hspace{1cm} T_3=L_{89} & \hspace{1cm} T_3^2 ~. 
\ea
The Casimir operators satisfy the relations \cite{Barut}
\ba
T^2 &=& \Lambda_6^2 + \frac{15}{4} ~, 
\nonumber\\
T_3^2 &=& \Lambda_7^2 + \frac{25}{4} ~. 
\label{Casimir}
\ea
This indicates that there exists a complementarity relationship 
\cite{Quesne} between the groups $SO(2,1)$ and $SO(6)$ within the 
single irreducible representation of $SO(7,2)$. 
As a consequence, the irreducible representations 
of $SO(6)$ are determined by those of $SO(2,1)$. 
The same holds for $SO(7)$ and $SO(2)$. 
The basis states $|\tilde{\psi}\rangle$ 
can be classified either by the $SO(2,1) \supset SO(2)$ 
labels $q,q_0$ which span the discrete representation $D^+$ of $SO(2,1)$, 
or by the $SO(7) \supset SO(6)$ labels $\omega,\gamma$ which span 
the symmetric irreducible representation of $SO(7)$.   
They are simultaneous eigenfunctions of the Casimir operators 
of the subgroups in Eq.~(\ref{lattice}) and satisfy the eigenvalue 
equations 
\ba 
T^2 \, | \tilde{\psi} \rangle &=& q(q+1) \, | \tilde{\psi} \rangle ~, 
\hspace{2cm} (q \mbox{ real and }< 0 ) ~, 
\nonumber\\ 
T_3 \, | \tilde{\psi} \rangle &=& q_0 \, | \tilde{\psi} \rangle ~, 
\hspace{3cm} (q_0=-q+s ~, \mbox{ with } s=0,1,\ldots) ~, 
\nonumber\\ 
\Lambda_7^2 \, | \tilde{\psi} \rangle &=& \omega(\omega+5) \, 
| \tilde{\psi} \rangle ~, 
\hspace{2cm} (\omega=0,1,\ldots) ~,  
\nonumber\\ 
\Lambda_6^2 \, | \tilde{\psi} \rangle &=& \gamma(\gamma+4) \, 
| \tilde{\psi} \rangle ~, 
\hspace{2cm} (\gamma=0,1,\ldots,\omega) ~.  
\label{eveq}
\ea 
Combining Eqs.~(\ref{Casimir}) and (\ref{eveq}), we find 
\ba
q(q+1) &=& \gamma(\gamma+4) + \frac{15}{4} ~,
\nonumber\\
q_0^2 &=& \omega(\omega+5) + \frac{25}{4} ~, 
\ea
which, solving for $q<0$ and $q_0 \geq -q$, gives 
\ba
q \;=\; -\gamma - \frac{5}{2} ~, \hspace{1cm}
q_0 \;=\; \omega + \frac{5}{2} ~. \label{comp}
\ea

\section{Basis states}

For three identical particles, the hypercoulomb Hamiltonian of 
Eq.~(\ref{hamilt}) is invariant under the permutation group $S_3$, 
and hence its eigenstates also carry good permutation symmetry 
$t=S$, $M$ or $A$, corresponding to the symmetric, mixed symmetric, 
and antisymmetric irreducible representation of $S_3$, respectively. 
In order to incorporate the permutation symmetry we 
associate the coordinates $r_j$ with the 
Jacobi vectors for the three-body problem \cite{KM} 
\ba
\vec{\rho} &=& \frac{1}{\sqrt{2}} (\vec{x}_1 - \vec{x}_2) 
\;\equiv\; (r_1,r_2,r_3) ~,
\nonumber\\
\vec{\lambda} &=& \frac{1}{\sqrt{6}} (\vec{x}_1 + \vec{x}_2 -2\vec{x}_3) 
\;\equiv\; (r_4,r_5,r_6) ~.
\label{jacobi}
\ea
Here $\vec{x}_1$, $\vec{x}_2$ and $\vec{x}_3$ denote the coordinates 
of the three particles. 
A convenient set of basis states is provided by the 
irreducible representations of the group chain 
\ba
\left| \begin{array}{ccccccccccc} 
SO(7) &\supset& SO(6) &\supset& 
[SU(3) &\supset& SO(3) &\supset& SO(2)] &\otimes& SO(2) \\ 
\omega &,& \gamma &,& (n_1,n_2) &,& L &,& M_L &,& \nu 
\end{array} \right> ~. \label{ch2} 
\ea
The reduction of $SO(6)$ can be obtained by using the complementarity 
relationship between the groups $SU(3)$ and $SO(2)$ within the symmetric 
irreducible representation $SO(6)$ \cite{Chacon}. 
As a consequence, the labels of $SU(3)$ are determined by that of $SO(2)$. 
The branching rules are given by
\ba
\gamma &=& 0,1,\ldots,\omega ~, 
\nonumber\\
\nu &=& -\gamma,-\gamma+2,\ldots,\gamma ~, 
\nonumber\\
(n_1,n_2) &=& (\gamma,\frac{\gamma-\nu}{2}) ~. \label{br2}
\ea
The reduction from $SU(3)$ to the rotation group $SO(3)$ 
is not fully reducible. In order to label the states uniquely 
an extra label is needed. We follow the procedure of \cite{Elliott} 
and introduce instead of $(n_1,n_2)$ 
the labels $(\lambda,\mu)=(n_1-n_2,n_2)=((\gamma+\nu)/2,(\gamma-\nu)/2)$~.
The values of $L$ contained in $(\lambda,\mu)$ are given by 
\cite{Elliott}
\ba
\kappa &=& \mbox{min}\{\lambda,\mu\}, 
\mbox{min}\{\lambda,\mu\}-2,\ldots,1 \mbox{ or } 0 ~, 
\nonumber\\
L &=& \left\{ \begin{array}{ll} \mbox{max}\{\lambda,\mu\},
\mbox{max}\{\lambda,\mu\}-2,\ldots,1 \mbox{ or } 0 ~, & 
\hspace{1cm} \mbox{for } \kappa \;=\; 0 ~, \\
\kappa,\kappa+1,\ldots,\kappa+\mbox{max}\{\lambda,\mu\} ~, &
\hspace{1cm} \mbox{for } \kappa \;>\; 0 ~, \end{array} \right.
\nonumber\\
M_L &=& -L,-L+1,\ldots,L ~. 
\ea
The $S_3$ invariant states are given by the linear combinations 
\ba
\frac{-i}{\sqrt{2(1+\delta_{\nu,0})}} \, 
\left( |  \nu \rangle - | -\nu \rangle \right) ~,
\nonumber\\
\frac{(-1)^{\nu_0}}{\sqrt{2(1+\delta_{\nu,0})}} \, 
\left( |  \nu \rangle + | -\nu \rangle \right) ~. 
\label{wfp12}
\ea
Here we have introduced the label $\nu_0$ by $\nu=\nu_0$ (mod 3). 
These wave functions transform for $\nu_0=0$ as $t=A$, $S$, and 
for $\nu_0=1,2$ as the two components of the mixed symmetric 
representation $t=M$. 
Summarizing, the basis states of the hypercoulomb problem can 
be characterized uniquely by 
\ba
| \tilde{\psi} \rangle &=& 
| \, \omega,\gamma,|\nu|,\kappa,L^P_t,M_L \rangle ~. \label{basis2}
\ea
Here $P=(-)^{\gamma}$ denotes the parity.
In Table~\ref{basis} we present the classification scheme of the 
basis states for $\omega=4$. 

\section{Energy spectrum} 

The energy spectrum of the hypercoulomb problem in six dimensions 
can be obtained by using the properties of the $SO(2,1)$ spectrum 
generating algebra \cite{Barut,Wybourne}. The Schr\"odinger equation 
can be expressed in terms of the generators of 
$SO(2,1)$ by introducing 
\ba
{\cal O} \, | \psi \rangle \;\equiv\; 
r(H-E) \, | \psi \rangle \;=\; 0 ~,
\ea
with
\ba
{\cal O} &=& ( \frac{1}{2\mu}-E ) \, T_3 + ( \frac{1}{2\mu}+E ) \, T_1 
- \tau ~.
\ea
This equation can be simplified further by performing a rotation about a
tilting angle $\theta$ 
\ba
\tilde{\cal O} \, | \tilde{\psi} \rangle &=& 0 ~, \label{seq1}
\ea
with
\ba
| \tilde{\psi} \rangle &=& {\cal N} \, \mbox{e}^{-i \theta T_2} \, 
| \psi \rangle ~,
\nonumber\\
\tilde{\cal O} &=& \mbox{e}^{-i \theta T_2} \, {\cal O} \, 
\mbox{e}^{i \theta T_2} 
\nonumber\\
&=& ( \frac{1}{2\mu}-E ) \, (T_3 \cosh \theta + T_1 \sinh \theta) 
  + ( \frac{1}{2\mu}+E ) \, (T_1 \cosh \theta + T_3 \sinh \theta) 
- \tau ~. 
\label{seq2}
\ea
The tilting angle may be chosen to diagonalize either the compact 
generator $T_3$ for the bound states, or the noncompact 
generator $T_1$ for the continuous states. The discrete spectrum 
is obtained by the choice
\ba
\tanh \theta &=& \frac{E+\frac{1}{2\mu}}{E-\frac{1}{2\mu}} ~, 
\ea
which reduces Eqs.~(\ref{seq1}) and (\ref{seq2}) 
to an eigenvalue equation of $T_3$
\ba
\left( \sqrt{-2E/\mu} \, T_3 - \tau \right) \, | \tilde{\psi} \rangle &=& 0 ~.
\label{seq}
\ea
The energy eigenvalues of the hypercoulomb problem are 
obtained by solving Eq.~(\ref{seq}) 
\ba
E \;=\; - \frac{\mu\tau^2}{2n^2} ~, \hspace{2cm} 
(n \;=\; q_0 \;=\; \omega + \frac{5}{2}) ~. 
\ea
The corresponding tilting angle is given by
\ba
\theta &=& - \ln (n/\mu\tau) ~.
\ea
The physical eigenstates, {\it i.e.} the solutions of the Schr\"odinger 
equation for the hypercoulomb problem, are related to the group states 
by a tilting rotation 
\ba
| \psi \rangle &=& \frac{1}{{\cal N}} \, 
\mbox{e}^{-i \ln(n/\mu\tau) \, T_2} \, 
| \tilde{\psi} \rangle ~.
\ea
They satisfy the normalization condition (note that the group 
metric is $1/r$ \cite{BR}) 
\ba
1 \;=\; \langle \psi | \psi \rangle &=& \frac{1}{{\cal N}^2} 
\langle \tilde{\psi} | \, \mbox{e}^{i \ln(n/\mu\tau) \, T_2} \, r \, 
\mbox{e}^{-i \ln (n/\mu\tau) \, T_2} \, | \tilde{\psi} \rangle 
\nonumber\\
&=& \frac{1}{{\cal N}^2} \, 
\langle \tilde{\psi} | \, \mbox{e}^{i \ln(n/\mu\tau) \, T_2} \, (T_3-T_1) \, 
\mbox{e}^{-i \ln(n/\mu\tau) \, T_2} \, | \tilde{\psi} \rangle 
\nonumber\\
&=& \frac{1}{{\cal N}^2 \mu\tau} \, 
\langle \tilde{\psi} | \, n(T_3-T_1) \, | \tilde{\psi} \rangle 
\nonumber\\
&=& \frac{n^2}{{\cal N}^2 \mu\tau} ~. \label{norm}
\ea
Hence, the normalized physical eigenstates can be expressed in 
terms of the group states as 
\ba
| \psi \rangle &=& \frac{\sqrt{\mu\tau}}{n} \, 
\mbox{e}^{-i \ln(n/\mu\tau) \, T_2} \, 
| \tilde{\psi} \rangle ~. \label{wf}
\ea

\section{Wave functions}

The wave function $| \tilde{\psi} \rangle$ satisfies the eigenvalue 
equations of Eq.~(\ref{eveq}), and hence is a $SO(2,1)$ 
eigenstate $| \tilde{\psi} \rangle = | q,q_0 \rangle$ with 
\ba
| q,q_0 \rangle &=&  
\sqrt{\frac{(-2q-1)!}{(q_0+q)!(q_0-q-1)!}} \, 
( T_+ )^{q_0+q} \, | q,-q \rangle ~,
\nonumber\\
\langle q,q_0 | &=& 
\sqrt{\frac{(-2q-1)!}{(q_0+q)!(q_0-q-1)!}} \, 
\langle q,-q | \, ( T_- )^{q_0+q} ~, \label{qq0}
\ea
with $T_{\pm}=T_1 \pm i T_2$. 
The action of $T_3$ and the raising and lowering operators 
$T_{\pm}$ on the $SO(2,1)$ 
eigenstates $| q,q_0 \rangle$ is given by \cite{HB} 
\ba 
T_3 \, | q,q_0 \rangle &=& q_0 \, | q,q_0 \rangle ~,
\nonumber\\
T_{\pm} \, | q,q_0 \rangle &=& \sqrt{-(q \mp q_0)(q \pm q_0 + 1 )} \, 
| q,q_0 \pm 1 \rangle ~.
\ea
The ground state of the hypercoulomb potential has $\omega=0$, and  
hence the principal quantum number is $n=q_0=\omega+5/2=5/2$. 
Moreover, since the only allowed value of $\gamma$ is $\gamma=0$, 
we have $q=-\gamma-5/2=-5/2$. 
Excited states with the same value of $q$, but 
with different values of $q_0$, can be created by applying the step 
operators $T_{\pm}$ according to Eq.~(\ref{qq0}). 
Since $T_{\pm}$ commute with the generators of $SO(6)$ (and its subgroups), 
these operators do not change the quantum numbers associated with the 
irreducible representations of $SO(6)$ and its subgroups $\gamma$, 
$|\nu|$, $\kappa$, $L^P_t$ and $M_L$. The operators $T_{\pm}$ are step 
operators in $q_0$, and hence change the value of the $SO(7)$ label 
$\omega$ by one unit. 
Excited states with the same value of $q_0$, but 
with different values of $q$, can be created by using 
the operators $L_{j7}$ with $j=1,\ldots,6$. 
Since $L_{j7}$ is a generator of $SO(7)$, it does not change the 
value of $\omega$. However, under $SO(6)$ it transforms as a 
six-dimensional vector, and hence changes the value of $\gamma$ by 
one unit. The transformation properties of $T_{\pm}$ and 
$L_{j7}$ under $SO(2,1)$ give rise to the selection rules
\ba
\begin{array}{rll}
T_{\pm} \, : & \hspace{1cm} \Delta q   \;=\; \Delta \gamma \;=\; 0 ~, 
        & \hspace{1cm} \Delta q_0 \;=\; \Delta \omega \;=\; \pm 1 ~, \\
L_{j7} \, : & \hspace{1cm} \Delta q   \;=\; \Delta \gamma \;=\; \pm 1 ~, 
        & \hspace{1cm} \Delta q_0 \;=\; \Delta \omega \;=\; 0 ~. 
\end{array}
\ea
Summarizing, the $SO(2,1)$ states 
$| \tilde{\psi} \rangle = | q,q_0 \rangle$ can be created 
from the ground state by applying the operators $T_+$ and $L_{j7}$
\ba
\begin{array}{ccccccc}
| -\frac{5}{2},\frac{5}{2} \rangle &\stackrel{T_+}{\longrightarrow}& 
| -\frac{5}{2},\frac{7}{2} \rangle &\stackrel{T_+}{\longrightarrow}& 
| -\frac{5}{2},\frac{9}{2} \rangle &\stackrel{T_+}{\longrightarrow}& 
\ldots \\
& &\updownarrow \, L_{j7}& &\updownarrow \, L_{j7} & & \\
& & | -\frac{7}{2},\frac{7}{2} \rangle &\stackrel{T_+}{\longrightarrow}& 
| -\frac{7}{2},\frac{9}{2} \rangle &\stackrel{T_+}{\longrightarrow}& 
\ldots \\
& & & &\updownarrow \, L_{j7} & & \\
& & & & | -\frac{9}{2},\frac{9}{2} \rangle 
&\stackrel{T_+}{\longrightarrow}& \ldots 
\end{array} \label{so21wf}
\ea
Similarly, the ket states $\langle \tilde{\psi} | = \langle q,q_0 |$ 
can be created by using the operators $T_-$ and $L_{j7}$. 
It is important to note, that the $SO(7)$ wave functions 
$| \tilde{\psi} \rangle = | q,q_0 \rangle$ of Eq.~(\ref{qq0}) 
satisfy a different normalization condition, see Eq.~(\ref{norm}), 
than the $SO(7)$ wave functions of \cite{rb1,rb2}. 
The hypercoulomb wave functions, {\it i.e} the eigenfunctions 
$| \psi \rangle$ of the original Hamiltonian of Eq.~(\ref{hamilt}), 
correspond to the group states $| \tilde{\psi} \rangle$ 
tilted about $T_2$ by an angle $\theta=-\ln(n/\mu\tau)$, see Eq.~(\ref{wf}). 
Since $T_2$ commutes with the generators of $SO(6)$ (and its subgroups), 
the hypercoulomb wave function $| \psi \rangle$ still has good  
$\gamma$, $|\nu|$, $\kappa$, $L^P_t$ and $M_L$, but the $SO(7)$ 
label $\omega$ no longer is a good quantum number. 

\section{Oscillator realization} 

The oscillator realization, that is frequently used for hydrogenlike 
problems characterized by the dynamical group $SO(4,2)$, is based on 
the isomorphism between $SO(4)$ and $SU(2) \otimes SU(2)$, and the 
Schwinger realization of $SU(2)$ \cite{Hatom}. 
However, this realization cannot be generalized to other dimensions,
since it is based on a particular property of the three-dimensional case. 
In \cite{Kyr} an oscillator realization was proposed for the general 
$m-$dimensional case which is characterized by the dynamical group 
$SO(m+1,2)$. For six dimensions ($m=6$) this gives a realization of 
the algebra of $SO(7,2)$ in terms of seven boson creation and 
annihilation operators 
\ba
L_{jk} &=& -i ( a^{\dagger}_j a_k - a^{\dagger}_k a_j ) ~, 
\nonumber\\
L_{j8} &=& \frac{1}{2} i ( \sum_{k=1}^7 a^{\dagger}_k a^{\dagger}_k a_j 
-2 \sum_{k=1}^7 a^{\dagger}_j a^{\dagger}_k a_k 
-5 a^{\dagger}_j + a_j ) ~,
\nonumber\\
L_{j9} &=& -\frac{1}{2} ( \sum_{k=1}^7 a^{\dagger}_k a^{\dagger}_k a_j 
-2 \sum_{k=1}^7 a^{\dagger}_j a^{\dagger}_k a_k 
-5 a^{\dagger}_j - a_j ) ~,
\nonumber\\
L_{89} &=& \sum_{k=1}^7 a^{\dagger}_k a_k + \frac{5}{2} ~, 
\hspace{2cm} (j,k=1,\ldots,7) ~. 
\label{osc}
\ea
The basis states of Eq.~(\ref{basis2}) correspond to linear combinations 
of the oscillator states 
\ba
\prod_{i=1}^7 \frac{1}{\sqrt{n_i !}} 
( a_i^{\dagger} )^{n_i} \, | 0 \rangle ~, 
\ea
where the total number of oscillator quanta 
is $\sum_i n_i = \omega$. 
According to Eq.~(\ref{jacobi}), we associate $a^{\dagger}_j$ with 
$j=1,\ldots,6$ with the components of the two Jacobi vector boson 
operators and $a^{\dagger}_7$ with a scalar boson operator 
\ba 
a^{\dagger}_k \;\equiv\; p^{\dagger}_{\rho,k} ~, \hspace{1cm} 
a^{\dagger}_{k+3} \;\equiv\; p^{\dagger}_{\lambda,k} ~, \hspace{1cm} 
a^{\dagger}_7 \;\equiv\; s^{\dagger} ~,
\ea
with $k=1,2,3$. The oscillator realization of Eq.~(\ref{osc}) 
is especially useful in deriving matrix elements.  

\section{Form factors}

The six-dimensional hypercoulomb potential is an exactly 
solvable model, and hence provides a set of closed 
expressions for the spectrum and the form factors . 
The hypercoulomb ground state $| \psi_0 \rangle = \sqrt{\mu\tau} \, 
\mbox{e}^{-i \ln(n_0/\mu\tau) \, T_2} \, | \tilde{\psi}_0 \rangle/n_0$ 
is characterized by the principal quantum number $n_0=5/2$. 
According to the oscillator realization of Eq.~(\ref{osc}), 
the ground state has zero oscillator quanta 
\ba
| \tilde{\psi}_0 \rangle \;\equiv\; | 0 \rangle ~, \hspace{2cm} 
\langle \tilde{\psi}_0 | \;\equiv\; \langle 0 | ~.
\label{gs}
\ea
Excited states can be created by the procedure outlined in Section~5 
in Eqs.~(\ref{qq0}) and (\ref{so21wf}). The oscillator realization of 
some $SO(2,1)$ eigenstates $| q,q_0 \rangle$ and $\langle q,q_0 |$ 
is given in Table~\ref{wfosc}. 

The derivation of the elastic form factor requires the evaluation 
of the matrix element of $\mbox{exp}(i k \lambda_z)=\mbox{exp}(i k r_6)$ 
in the ground state $| \psi_0 \rangle$ 
\ba
F_0(k) &=& \langle \psi_0 | \, \mbox{e}^{i k r_6} \, | \psi_0 \rangle 
\nonumber\\
&=& \frac{\mu\tau}{n_0^2} \, \langle \tilde{\psi}_0 | \, 
\mbox{e}^{ i \ln(n_0/\mu\tau) \, T_2} \, r \, \mbox{e}^{i k r_6} \, 
\mbox{e}^{-i \ln(n_0/\mu\tau) \, T_2} \, | \tilde{\psi}_0 \rangle 
\ea 
According to Eq.(\ref{gen1}), both the coordinate $r_j$ 
and the radius $r$ can be expressed in terms of a linear 
combination of generators 
\ba
r_j \;=\; L_{j8}-L_{j7} ~, \hspace{1cm} 
r \;=\; L_{89}-L_{79} ~.
\ea
This allows one to express $F_0(k)$ in terms of the generators 
of $SO(7,2)$ 
\ba
F_0(k) &=& \frac{\mu\tau}{n_0^2} \, \langle \tilde{\psi}_0 | \, 
\mbox{e}^{ i \ln(n_0/\mu\tau) \, L_{78}} \, (L_{89}-L_{79}) \, 
\mbox{e}^{i k (L_{68}-L_{67})}  \, 
\mbox{e}^{-i \ln(n_0/\mu\tau) \, L_{78}} \, | \tilde{\psi}_0 \rangle 
\nonumber\\
&=& \frac{1}{n_0^2} \, \langle \tilde{\psi}_0 | \, 
n_0(L_{89}-L_{79}) \, \mbox{e}^{i kn_0 (L_{68}-L_{67})/\mu\tau} \, 
| \tilde{\psi}_0 \rangle 
\nonumber\\
&=& \langle \tilde{\psi}_0 | \, \mbox{e}^{i kn_0 (L_{68}-L_{67})/\mu\tau} \, 
| \tilde{\psi}_0 \rangle - \frac{1}{n_0} \, \langle \tilde{\psi}_0 | \, 
L_{79} \, \mbox{e}^{i kn_0 (L_{68}-L_{67})/\mu\tau} \, 
| \tilde{\psi}_0 \rangle ~. 
\label{f0k}
\ea
(i) The first term in Eq.~(\ref{f0k}) can be evaluated by rewriting 
\ba
\mbox{e}^{i kn_0 (L_{68}-L_{67})/\mu\tau} &=& \mbox{e}^{-i \alpha L_{67}} \,
\mbox{e}^{-i \beta L_{78}} \, \mbox{e}^{-i \gamma L_{67}} ~, 
\label{top1}
\ea
with 
\ba
\sinh (\beta/2) &=& \pm kn_0/2\mu\tau ~, 
\nonumber\\
\cosh (\beta/2) &=& \left( 1 + (kn_0/2\mu\tau)^2 \right)^{1/2} ~,
\nonumber\\
\sin \alpha &=& - \sin \gamma \;=\; \pm \frac{1}{\cosh (\beta/2)} ~,
\nonumber\\
\cos \alpha &=& - \cos \gamma \;=\; - \tanh (\beta/2) ~. 
\label{angles}
\ea
Since the ground state has zero oscillator quanta, the matrix 
element of Eq.~(\ref{top1}) reduces to 
\ba
\langle \tilde{\psi}_0 | \, \mbox{e}^{i kn_0 (L_{68}-L_{67})/\mu\tau} \, 
| \tilde{\psi}_0 \rangle &=& 
\langle \tilde{\psi}_0 | \, \mbox{e}^{-i \alpha L_{67}} \,
\mbox{e}^{-i \beta L_{78}} \, \mbox{e}^{-i \gamma L_{67}} \, 
| \tilde{\psi}_0 \rangle ~,
\nonumber\\
&=& \langle \tilde{\psi}_0 | \, \mbox{e}^{-i \beta L_{78}} \, 
| \tilde{\psi}_0 \rangle ~. \label{fk1}
\ea
The ground state wave function satisfies 
\ba
T_3 \, | \tilde{\psi}_0 \rangle &=& n_0 \, | \tilde{\psi}_0 \rangle ~,
\nonumber\\
T_- \, | \tilde{\psi}_0 \rangle &=& 
a_7 \, | \tilde{\psi}_0 \rangle \;=\; 0 ~,
\ea
and can be characterized by $|q=-n_0,q_0=n_0 \rangle$. The ground state 
expectation value of $\mbox{exp}(-i \beta L_{78})$ in Eq.~(\ref{fk1})  
is a representation matrix element 
for the discrete series $D^+$ of $SO(2,1)$ \cite{HB} 
\ba
\langle \tilde{\psi}_0 | \, \mbox{e}^{-i \beta L_{78}} \, 
| \tilde{\psi}_0 \rangle \;=\; 
\langle -n_0,n_0 | \, \mbox{e}^{-i \beta T_2} \, | -n_0,n_0 \rangle 
\;=\; \left( \cosh \frac{\beta}{2} \right)^{-2n_0} ~. 
\label{term1}
\ea
(ii) The second contribution to the form factor 
of Eq.~(\ref{f0k}) can be rewritten as 
\ba
\langle \tilde{\psi}_0 | \, 
L_{79} \, \mbox{e}^{i kn_0 (L_{68}-L_{67})/\mu\tau} \, 
| \tilde{\psi}_0 \rangle 
&=& \langle \tilde{\psi}_0 | \, L_{79} \, \mbox{e}^{-i \alpha L_{67}} \, 
\mbox{e}^{-i \beta L_{78}} \, \mbox{e}^{-i \gamma L_{67}} \, 
| \tilde{\psi}_0 \rangle 
\nonumber\\
&=& \langle \tilde{\psi}_0 | \,( L_{79} \cos \alpha + L_{69} \sin \alpha ) \, 
\mbox{e}^{-i \beta L_{78}} \, | \tilde{\psi}_0 \rangle ~. \label{sine}
\ea
The first term on the right hand side can be expressed in terms of the 
derivative of Eq.~(\ref{term1}) 
\ba
\langle \tilde{\psi}_0 | \, L_{79} \, 
\mbox{e}^{-i \beta L_{78}} \, | \tilde{\psi}_0 \rangle 
&=& -i \, \langle \tilde{\psi}_0 | \, L_{78} \, 
\mbox{e}^{-i \beta L_{78}} \, | \tilde{\psi}_0 \rangle 
\nonumber\\
&=& \frac{d}{d \beta} \, \langle \tilde{\psi}_0 | \, 
\mbox{e}^{-i \beta L_{78}} \, | \tilde{\psi}_0 \rangle
\nonumber\\ 
&=& - n_0 \, \left( \sinh \frac{\beta}{2} \right) \, 
\left( \cosh \frac{\beta}{2} \right)^{-2n_0-1} ~. 
\label{term2a} 
\ea
The second term in Eq.~(\ref{sine}), proportional to $\sin \alpha$, 
vanishes identically. 
This can be seen by introducing the $SO(2,1) \otimes SO(2,1)$ algebra 
\ba
\left[ G_1,G_2 \right] \;=\; -i G_3 ~, \hspace{1cm} 
&\left[ G_2,G_3 \right] \;=\;  i G_1 ~,& \hspace{1cm} 
\left[ G_3,G_1 \right] \;=\;  i G_2 ~,
\nonumber\\
\left[ H_1,H_2 \right] \;=\; -i H_3 ~, \hspace{1cm} 
&\left[ H_2,H_3 \right] \;=\;  i H_1 ~,& \hspace{1cm} 
\left[ H_3,H_1 \right] \;=\;  i H_2 ~,
\nonumber\\
&\left[ G_i,H_j \right] \;=\; 0 ~,&
\ea
which is generated by the operators \cite{Kyr} 
\ba
G_1 \;=\; \frac{1}{2} ( L_{68} + L_{79} ) ~, &\hspace{1cm}& 
H_1 \;=\; \frac{1}{2} ( L_{68} - L_{79} ) ~,
\nonumber\\
G_2 \;=\; \frac{1}{2} ( L_{78} - L_{69} ) ~, &\hspace{1cm}& 
H_2 \;=\; \frac{1}{2} ( L_{78} + L_{69} ) ~,
\nonumber\\
G_3 \;=\; \frac{1}{2} ( L_{67} + L_{89} ) ~, &\hspace{1cm}& 
H_3 \;=\; \frac{1}{2} ( L_{67} - L_{89} ) ~.
\ea
We rewrite the second term in terms of the operators $G_i$ and $H_j$ 
\ba
\langle \tilde{\psi}_0 | \, L_{69} \, 
\mbox{e}^{-i \beta L_{78}} \, | \tilde{\psi}_0 \rangle 
&=& \langle \tilde{\psi}_0 | \, (H_2-G_2) \, 
\mbox{e}^{-i \beta (H_2+G_2)} \, | \tilde{\psi}_0 \rangle 
\nonumber\\
&=& \langle \tilde{\psi}_0 | \, H_2 \, \mbox{e}^{-i \beta H_2} \, 
\mbox{e}^{-i \beta G_2} \, | \tilde{\psi}_0 \rangle 
- \langle \tilde{\psi}_0 | \, G_2 \, \mbox{e}^{-i \beta G_2} \, 
\mbox{e}^{-i \beta H_2} \, | \tilde{\psi}_0 \rangle ~. \label{gh}
\ea
The action of $G_3$ and $H_3$ and the raising and lowering operators 
$G_{\pm} = G_1 \pm i G_2$ and $H_{\pm} = H_1 \pm i H_2$ 
on the ground state wave function can be evaluated by 
using the oscillator realization of Eq.~(\ref{osc}) and  
Eq.~(\ref{gs}) 
\ba
G_3 \, | \tilde{\psi}_0 \rangle \;=\; 
- H_3 \, | \tilde{\psi}_0 \rangle &=& 
\frac{n_0}{2} \, | \tilde{\psi}_0 \rangle ~,
\nonumber\\
G_- \, | \tilde{\psi}_0 \rangle \;=\; 
H_+ \, | \tilde{\psi}_0 \rangle &=& 0 ~. \label{ext}
\ea
Thus, the ground state wave function is characterized by 
$g=-g_0=-n_0/2$ and $h=h_0=-n_0/2$. Eq.~(\ref{ext}) shows that 
the matrix element of $\mbox{exp}(-i \beta G_2)$ is a 
representation matrix element for the positive discrete 
series $D^+$, whereas that of $\mbox{exp}(-i \beta H_2)$ 
corresponds to a representation matrix element for the 
negative discrete series $D^-$. The relevant matrix elements 
are given by \cite{HB} 
\ba
\left< -\frac{n_0}{2}, \frac{n_0}{2} \right| \, \mbox{e}^{-i \beta G_2} \, 
\left| -\frac{n_0}{2}, \frac{n_0}{2} \right> \;=\; 
\left< -\frac{n_0}{2},-\frac{n_0}{2} \right| \, \mbox{e}^{-i \beta H_2} \, 
\left| -\frac{n_0}{2},-\frac{n_0}{2} \right> 
\;=\; \left( \cosh \frac{\beta}{2} \right)^{-n_0} ~. 
\ea
Therefore, the two terms on the right hand side of Eq.~(\ref{gh}) 
cancel, and the matrix element 
\ba
\langle \tilde{\psi}_0 | \, L_{69} \, 
\mbox{e}^{-i \beta L_{78}} \, | \tilde{\psi}_0 \rangle &=& 0 ~, 
\label{term2b}
\ea
vanishes identically.

Summarizing, the elastic form factor of Eq.~(\ref{f0k}) is obtained by 
combining the contributions from Eqs.~(\ref{term1}), (\ref{term2a}) 
and (\ref{term2b})
\ba
F_0(k) &=& \left( \cosh \frac{\beta}{2} \right)^{-2n_0-2}
\;=\; \frac{1}{\left( 1 + \frac{25}{16\mu^2\tau^2}k^2 \right)^{7/2}} ~, 
\label{result}
\ea
where we have used Eq.~(\ref{angles}) and $n_0=5/2$ for the ground state 
wave function. This result is in agreement with \cite{SIG,Leal}, 
in which the form factor was derived as an integral in bispherical 
coordinates. The derivation of transition form factors connecting 
the ground state to excited states proceeds in a similar way. 
The results are given in Table~\ref{ff}. 

The general result for the elastic form factor of the $m$-dimensional 
hypercoulomb potential can be obtained by using the properties of 
$SO(m+1,2)$ dynamical group. The derivation proceeds along the same 
lines as used in this section for $m=6$. The elastic form factor 
is given by 
\ba
F_0(k) &=& \frac{1}{\left( 1 + \frac{(m-1)^2}{16\mu^2\tau^2}k^2 
\right)^{(m+1)/2}} ~, 
\ea
and shows a power-law dependence on $k$. 
For the six-dimensional case it reduces to Eq.~(\ref{result}). 

\section{Nucleon form factors}

The hypercoulomb interaction whose properties have been described in 
the previous sections, can be used to analyze, at least approximately, 
nucleon phenomenology. In order to do so, one has first to 
assign the observed nucleon resonances to states of the hypercoulomb 
potential. This is done in Table~\ref{baryons}. 
The spatial wave function has to be combined 
with the spin-flavor and color parts, in such a way that the total 
wave function is antisymmetric (see {\it e.g.} \cite{SIG,IK,BIL}). 
The nucleon itself is identified with the ground state 
$| \tilde{\psi} \rangle = | 0,0,0^+_S \rangle$ which has $n=5/2$, 
the Roper resonance is associated with 
$| \tilde{\psi} \rangle = | 1,0,0^+_S \rangle$ and $n=7/2$, 
and the negative parity resonances with 
$| \tilde{\psi} \rangle = | 1,1,1^-_M \rangle$ and $n=7/2$.

\begin{figure}
\centerline{\hbox{
\psfig{figure=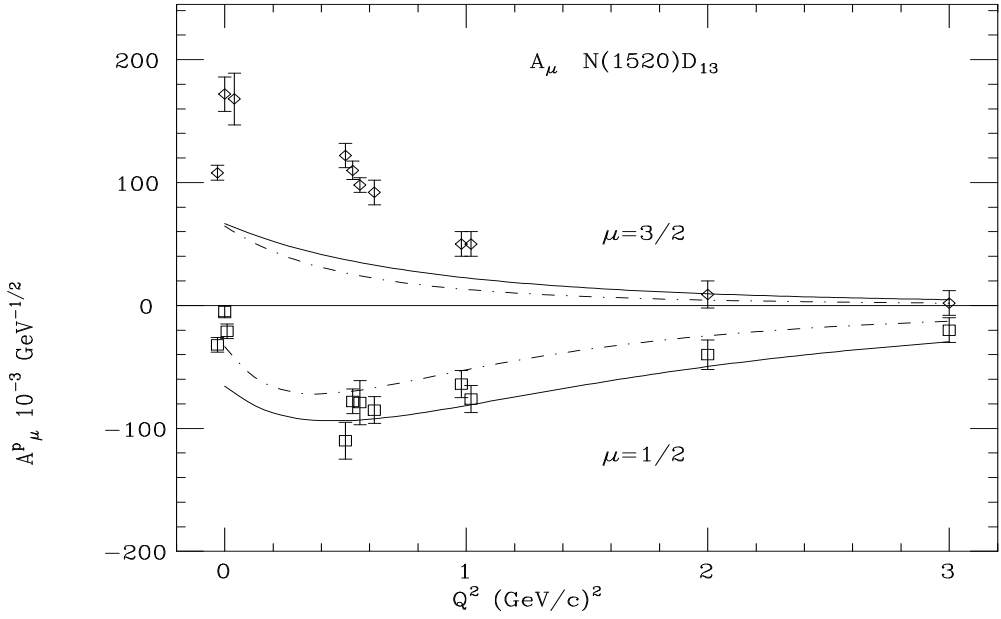,height=0.65\textwidth,width=1.0\textwidth,angle=90} }}
\caption[]{The helicity amplitudes $A^p_{1/2}$ and $A^p_{3/2}$ in the 
Breit frame for the excitation of the N(1520)$D_{13}$ resonance. 
The full curve is the complete result of the model of \protect\cite{FGPS}, 
while the dashed curve is the result without the linear confining term. 
The experimental data are taken from \protect\cite{data}.}
\label{d13}
\end{figure}

\begin{figure}
\centerline{\hbox{
\psfig{figure=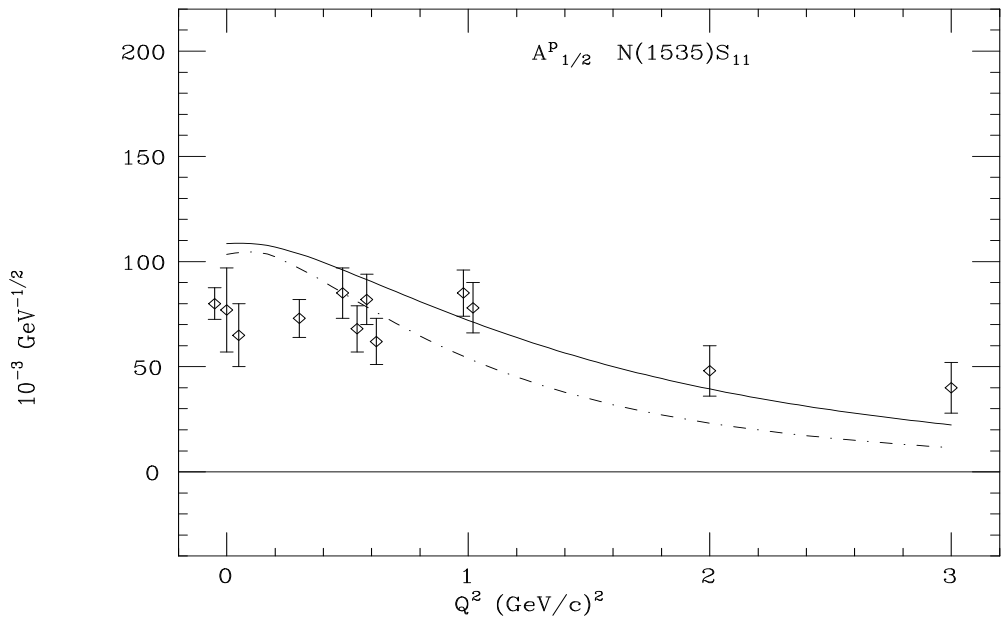,height=0.65\textwidth,width=1.0\textwidth,angle=90} }}
\caption[]{The same as in Fig.~\protect\ref{d13}, but for the excitation 
of the N(1535)$S_{11}$ resonance.}
\label{s11}
\end{figure}

The formulas derived in the previous section provide a way to 
calculate form factors and helicity amplitudes between the ground state 
and the baryon resonances (see {\it e.g.} \cite{FGPS,BIL,CKO}).  
As an example, we show in Figs.~\ref{d13} and~\ref{s11} 
the helicity amplitudes leading to the resonances 
N(1520)$D_{13}$ and N(1535)$S_{11}$.  
The hypercoulomb results are compared with experimental data and with 
complete calculations from  
\cite{FGPS} which take into account also the linear confining term.
It is seen that the hypercoulomb results are comparable in quality with 
those of the second model \cite{FGPS}.  Since the strength of
the parameter of the linear confining term is not too strong, 
the hypercoulomb 
form factors provide a good lowest order approximation.
We also comment on the fact that the hypercoulomb interaction 
produces form factors that decrease as the inverse of a power of the 
momentum transfer $k$, in  
contrast with the harmonic oscillator form factors (the other exactly 
solvable case) that drop too fast as a gaussian \cite{CKO}. 
For example, the elastic form factor $G_E^p(k)$ behaves as
\ba
G_E^p(k) &=& \left\{ \begin{array}{ll}
1/(1+k^2 a^2)^{7/2} \hspace{1cm} & \mbox{Hypercoulomb} \\ & \\
\mbox{exp}(-k^2 \beta^2 /6) & \mbox{Harmonic oscillator } \cite{CKO} 
\end{array} \right. \label{gep0}
\ea
where $a^2=25/24\mu^2\tau^2$ (see Table~\ref{ff}). 

\section{Conclusions}

In this paper we have presented a completely algebraic treatment 
of the six-dimensional hypercoulomb problem in the context of the 
dynamical group $SO(7,2)$. We have developed a systematic way 
to construct the wave functions using step operators in the 
$SO(7) \supset SO(6)$ labels $\omega$, $\gamma$. This makes it 
possible to derive elastic form factors and transition matrix elements 
connecting the ground state to excited states in closed analytic form 
in an entirely algebraic way. In the derivation we have used an 
oscillator realization of the algebra of $SO(7,2)$. This procedure can 
readily be extended to the hypercoulomb problem in any number of 
dimensions, as it has been shown for the elastic form factor. 
We discussed briefly the application to baryon phenomenology, 
in which it was found that the form factors drop as powers of the 
momentum transfer, as it is observed experimentally.

In conclusion, the present algebraic method provides an alternative 
to solving hypercoulomb problems with integro-differential methods. 

\section*{Acknowledgements}

This work is supported in part by DGAPA-UNAM under project IN101997 
(R.B.), in part by DOE, Grant No. DE-FG02-91ER40608 (F.I.), 
and in part by INFN (E.S.).


\clearpage

\begin{table}
\centering
\caption[]{\small
Basis states of $SO(7)$ with $\omega=4$. 
The corresponding $SO(2,1)$ labels are $q_0=\omega+5/2$ and 
$q=-\gamma-5/2$. 
\normalsize}
\label{basis}
\vspace{15pt}
\begin{tabular}{ccrcl} 
\hline
& & & & \\
$\omega$ & $\gamma$ & $\nu$ & $\kappa$ & $L^P_t$ \\
& & & & \\
\hline
& & & & \\
4 & 0 &       0 & 0 & $0^+_S$ \\
  & 1 & $\pm$ 1 & 0 & $1^-_M$ \\
  & 2 & $\pm$ 2 & 0 & $0^+_M,2^+_M$ \\
  &   &       0 & 1 & $1^+_A,2^+_S$ \\
  & 3 & $\pm$ 3 & 0 & $1^-_S,1^-_A,3^-_S,3^-_A$ \\
  &   & $\pm$ 1 & 1 & $1^-_M,2^-_M,3^-_M$ \\
  & 4 & $\pm$ 4 & 0 & $0^+_M,2^+_M,4^+_M$ \\
  &   & $\pm$ 2 & 1 & $1^+_M,2^+_M,3^+_M,4^+_M$ \\ 
  &   &       0 & 0 & $0^+_S,2^+_A$ \\
  &   &         & 2 & $2^+_S,3^+_A,4^+_S$ \\
& & & \\
\hline
\end{tabular}
\end{table}

\clearpage

\begin{table}
\centering
\caption[]{\small 
Oscillator realization for the wave functions 
$| \tilde{\psi} \rangle = | q,q_0 \rangle$ and 
$\langle \tilde{\psi} | = \langle q,q_0 |$ 
with $\omega \leq 2$. 
For these states the labels $\nu$ and $\kappa$ are 
redundant.
\normalsize}
\label{wfosc}
\vspace{15pt}
\begin{tabular}{ccrclcc} 
\hline
& & & & & & \\
$q$ & $q_0$ & $\omega$ & $\gamma$ & $L^P_t$ & $M_L$ & oscillator \\
& & & & & & \\
\hline
& & & & & & \\
$-\frac{5}{2}$ & $\frac{5}{2}$ & 0 & 0 & $0^+_S$ & 0 & $| 0 \rangle$ \\
$-\frac{5}{2}$ & $\frac{5}{2}$ & 0 & 0 & $0^+_S$ & 0 & $\langle 0 |$ \\
& & & & & & \\
$-\frac{5}{2}$ & $\frac{7}{2}$ & 1 & 0 & $0^+_S$ & 0 
& $\frac{1}{\sqrt{5}} \langle 0 | s$ \\
$-\frac{7}{2}$ & $\frac{7}{2}$ & 1 & 1 & $1^-_{M_{\rho}}$ & 1 
& $\frac{i}{\sqrt{5}} \langle 0 | p_{\rho,1}$ \\
$-\frac{7}{2}$ & $\frac{7}{2}$ & 1 & 1 & $1^-_{M_{\lambda}}$ & 1 
& $\frac{i}{\sqrt{5}} \langle 0 | p_{\lambda,1}$ \\
& & & & & & \\
$-\frac{5}{2}$ & $\frac{9}{2}$ & 2 & 0 & $0^+_S$ & 0 
& $\frac{1}{2\sqrt{15}} \langle 0 | s^2$ \\
$-\frac{7}{2}$ & $\frac{9}{2}$ & 2 & 1 & $1^-_{M_{\rho}}$ & 1 
& $\frac{i}{\sqrt{35}} \langle 0 | s p_{\rho,1}$ \\
$-\frac{7}{2}$ & $\frac{9}{2}$ & 2 & 1 & $1^-_{M_{\lambda}}$ & 1 
& $\frac{i}{\sqrt{35}} \langle 0 | s p_{\lambda,1}$ \\
$-\frac{9}{2}$ & $\frac{9}{2}$ & 2 & 2 & $2^+_S$ & 2  
& $\frac{1}{2\sqrt{35}} \langle 0 | [ p_{\rho,1} p_{\rho,1} 
+ p_{\lambda,1} p_{\lambda,1} ]$ \\
$-\frac{9}{2}$ & $\frac{9}{2}$ & 2 & 2 & $1^+_A$ & 1 
& $-\frac{1}{\sqrt{70}} \langle 0 | [ p_{\rho,1} p_{\lambda,0} 
- p_{\lambda,1} p_{\rho,0} ]$ \\
$-\frac{9}{2}$ & $\frac{9}{2}$ & 2 & 2 & $2^+_{M_{\rho}}$ & 2 
& $\frac{1}{\sqrt{35}} \langle 0 | p_{\rho,1} p_{\lambda,1}$ \\
$-\frac{9}{2}$ & $\frac{9}{2}$ & 2 & 2 & $2^+_{M_{\lambda}}$ & 2 
& $\frac{1}{2\sqrt{35}} \langle 0 | [ p_{\rho,1} p_{\rho,1} 
- p_{\lambda,1} p_{\lambda,1} ]$ \\
$-\frac{9}{2}$ & $\frac{9}{2}$ & 2 & 2 & $0^+_{M_{\rho}}$ & 0 
& $\frac{1}{\sqrt{105}} \langle 0 | p_{\rho} \cdot p_{\lambda}$ \\
$-\frac{9}{2}$ & $\frac{9}{2}$ & 2 & 2 & $0^+_{M_{\lambda}}$ & 0 
& $\frac{1}{2\sqrt{105}} \langle 0 | [ p_{\rho} \cdot p_{\rho} 
- p_{\lambda} \cdot p_{\lambda} ]$ \\
& & & & & \\
\hline
\end{tabular}
\end{table}

\clearpage

\begin{table}
\centering
\caption[]{\small 
Form factors for the hypercoulomb problem.  
The hypercoulomb wave functions are given by Eq.~(\ref{wf}) 
$| \psi \rangle = \sqrt{\mu\tau} \, \mbox{e}^{-i \ln(n/\mu\tau) \, T_2} \, 
| \tilde{\psi} \rangle/n$ 
with $| \tilde{\psi} \rangle = | \omega,\gamma,L^P_t \rangle$ 
and $n=\omega+5/2$.
The initial state is $| \tilde{\psi} \rangle = | 0,0,0^+_S \rangle$ 
and $a=n_0/\sqrt{6}\mu\tau=5/2\sqrt{6}\mu\tau$. \normalsize}
\label{ff}
\vspace{15pt}
\begin{tabular}{cc|c}
\hline
& & \\
$\langle \tilde{\psi}^{\prime} | = \langle \omega,\gamma,L^P_t |$ & & 
$\langle \psi^{\prime} | \mbox{e}^{i\sqrt{2/3} \, kr_6} | \psi \rangle$ \\
& & \\
\hline
& & \\
$\langle 0,0,0^+_S |$ & & $\frac{1}{(1+k^2a^2)^{7/2}}$ \\ 
& & \\
$\langle 1,1,1^-_M |$ & & $-i\sqrt{7}~(\frac{5}{6})^4~(\frac{7}{6})^4 \, 
\frac{ka}{(1+\frac{49}{36}k^2a^2)^{9/2}}$ \\ 
& & \\
$\langle 1,0,0^+_S |$ & & $\sqrt{7}~(\frac{5}{6})^4~(\frac{7}{6})^5 \, 
\frac{k^2a^2}{(1+\frac{49}{36}k^2a^2)^{9/2}}$ \\ 
& & \\
$\langle 2,2,2^+_S |$ & & $-\frac{\sqrt{21}}{\sqrt{2}}~(\frac{5}{7})^5~
(\frac{9}{7})^5~\frac{k^2a^2} {(1+\frac{81}{49}k^2a^2)^{11/2}}$ \\ 
& & \\
$\langle 2,2,2^+_M |$ & & $\frac{\sqrt{21}}{\sqrt{2}}~(\frac{5}{7})^5~
(\frac{9}{7})^5~\frac{k^2a^2}{(1+\frac{81}{49}k^2a^2)^{11/2}}$ \\ 
& & \\
$\langle 2,2,0^+_M |$ & & $-\frac{\sqrt{21}}{2}~(\frac{5}{7})^5~
(\frac{9}{7})^5~\frac{k^2a^2} {(1+\frac{81}{49}k^2a^2)^{11/2}}$
\\ 
& & \\
$\langle 2,1,1^-_M |$ & & $-i\frac{\sqrt{7}}{3}~(\frac{5}{7})^4~
(\frac{9}{7})^5~\frac{ka[1+\frac{486}{49}k^2a^2]} {(1+\frac{81}{49}k^2a^2)
^{11/2}}$ \\ 
& & \\
$\langle 2,0,0^+_S |$ & & $\frac{\sqrt{3}}{2}~(\frac{5}{7})^4~
(\frac{9}{7})^5~\frac{k^2a^2[1+\frac{486}{49}k^2a^2]} 
{(1+\frac{81}{49}k^2a^2)^{11/2}}$ \\ 
& & \\
\hline
\end{tabular}
\end{table}

\clearpage

\begin{table}
\centering
\caption[]{\small 
Identification of the states of Eq.~(\ref{wf}) and Table~\ref{basis} 
(the quantum numbers are $\omega$, $\gamma$, $L^P_t$) 
with the observed baryon resonances. 
The spin, $S$, and the total angular momentum and parity,
$J^{P}$, is also indicated.  The experimental masses are from 
\protect\cite{PDG}.}
\label{baryons}
\vspace{15pt}
\begin{tabular}{cccccccccc}
\hline
& & & & & & & & & \\ 
$\omega$~$\gamma$~$L^{P}$ & & $S$ & & $J^{P}$  & & & Baryon 
& Status & $M_{exp}$(MeV) \\ 
& & & & & & & & & \\ 
\hline
& & & & & & & & & \\ 
  $0$~$0$~$0^+$ & & 
$\frac{1}{2}$ & & $\frac{1}{2}^+$ & & & N(938)~$P_{11}$ & **** & 938\\ 
  $0$~$0$~$0^+$ & & 
$\frac{3}{2}$ & & $\frac{3}{2}^+$ & & &  
$\Delta (1232)~P_{33}$ & **** & 1232\\ 
 $1$~$0$~$0^+$ & & 
$\frac{1}{2}$ & & $\frac{1}{2}^+$ & & & 
N(1440)~$P_{11}$ & **** & 1440 \\ 
 $1$~$0$~$0^+$ & &
$\frac{1}{2}$ & & $\frac{3}{2}^+$ & & & 
$\Delta (1600)~P_{33}$ & *** & 1600 \\ 
 $1$~$1$~$1^-$ & &
$\frac{1}{2}$ & &  $\frac{1}{2}^-$ & & & N(1535)~$S_{11}$ & **** & 1535\\ 
 $1$~$1$~$1^-$ & &
$\frac{1}{2}$ & &  $\frac{3}{2}^-$ & & & N(1520)~$D_{13}$ & **** & 1520\\ 
 $1$~$1$~$1^-$ & &
$\frac{3}{2}$ & &  $\frac{1}{2}^-$ & & & N(1650)~$S_{11}$ & **** & 1650\\ 
  $1$~$1$~$1^-$ & &
$\frac{3}{2}$ & & $\frac{3}{2}^-$ & & & N(1700)~$D_{13}$ & *** & 1700\\ 
 $1$~$1$~$1^-$ & &
$\frac{3}{2}$ & & $\frac{5}{2}^-$ & & & N(1675)~$D_{15}$ & **** & 1675\\ 
 $1$~$1$~$1^-$ & 
& $\frac{1}{2}$ & & $\frac{1}{2}^-$ & & & 
$\Delta (1620)~S_{31}$ & **** & 1620 \\ 
 $1$~$1$~$1^-$ & 
& $\frac{1}{2}$ & & $\frac{3}{2}^-$ & & & 
$\Delta (1700)~D_{33}$ & **** & 1700 \\ 
 $2$~$0$~$0^+$ & &
$\frac{1}{2}$ & & $\frac{1}{2}^+$ & & & N(1710)~$P_{11}$ & *** & 1710\\ 
 $2$~$2$~$2^+$ & &
$\frac{1}{2}$ & & $\frac{3}{2}^+$ & & & N(1720)~$P_{13}$ & **** & 1720\\ 
 $2$~$2$~$2^+$ & &
$\frac{1}{2}$ & & $\frac{5}{2}^+$ & & &
 N(1680)~$F_{15}$ & **** & 1680\\  
 $2$~$2$~$2^+$ & 
& $\frac{3}{2}$ & & $\frac{1}{2}^+$ & & &  
$\Delta (1910)~P_{31}$ & **** & 1910\\ 
 $2$~$2$~$2^+$ & &
$\frac{3}{2}$ & & $\frac{3}{2}^+$ & & & 
$\Delta (1920)~P_{33}$ & *** & 1920\\ 
 $2$~$2$~$2^+$ & 
& $\frac{3}{2}$ & & $\frac{5}{2}^+$ & & & 
$\Delta (1905)~F_{35}$ & **** & 1905\\ 
 $2$~$2$~$2^+$ & 
& $\frac{3}{2}$ & & $\frac{7}{2}^+$ & & & 
$\Delta (1950)~F_{37}$ & **** & 1950\\ 
$2$~$1$~$1^-$ & &
$\frac{1}{2}$ & & $\frac{1}{2}^-$ & & & 
$\Delta (1900)~S_{31}$ & *** & 1900\\ 
& & & & & & & & & \\ 
\hline
\end{tabular}
\end{table}

\clearpage

\clearpage

\end{document}